\documentclass[preprint,12pt]{elsarticle}

\usepackage{amssymb}
\usepackage{amsmath}
\usepackage{float}
\usepackage{multirow}
\usepackage{algorithm}
\usepackage{algpseudocode}
\usepackage{gensymb}
\usepackage[hidelinks]{hyperref}
\usepackage{listings}
\usepackage{subcaption}
\usepackage{adjustbox}
\usepackage{bm}
\usepackage{xcolor}

\usepackage{caption}
\usepackage{graphicx}
\usepackage{latexsym}
\usepackage{times}
\usepackage[pagewise]{lineno}

\usepackage{fmtcount}
\usepackage{marvosym}
\usepackage{bm}
\usepackage{chemformula}
\usepackage{adjustbox}
\usepackage{epstopdf}

\newcommand\yloc{2.5}

\journal{}

\begin{document}
\sloppy
\begin{frontmatter}



\title{An asynchronous discontinuous Galerkin method for combustion simulations}


\author[inst1]{Aswin Kumar Arumugam}
\author[inst1]{Konduri Aditya\corref{cor1}}

\affiliation[inst1]{organization={Department of Computational and Data Sciences, Indian Institute of Science},
            city={Bengaluru},
            postcode={560012},
            state={KA},
            country={India}}

\cortext[cor1]{Corresponding author.}



\begin{abstract}
The discontinuous Galerkin (DG) method has been widely considered in recent years to develop scalable flow solvers for its ability to handle discontinuities, such as shocks and detonations, with greater accuracy and high arithmetic intensity. However, its scalability is severely affected by communication bottlenecks that arise from data movement and synchronization at extreme scales. Recently, an asynchronous discontinuous Galerkin (ADG) method was proposed to reduce communication overhead by either relaxing synchronization or avoiding communication between processing elements (PEs). The numerical properties of the ADG method were verified by solving simple one-dimensional partial differential equations. In this study, the application of the ADG method is extended to complex chemically reacting flows, particularly to evaluate its efficacy in capturing flame fronts and detonations. New asynchrony-tolerant weighted essentially non-oscillatory (AT-WENO) limiters are derived to accurately capture flow discontinuities with communication delays at the PE boundaries. The numerical accuracy of the ADG method is verified on the premixed spontaneous ignition of a fuel-lean mixture with detailed chemistry. The progression of the flame front is accurately represented by the ADG method, and the numerical errors incurred at the PE boundaries turn out to be insignificant. The efficacy of the AT-WENO limiters in capturing discontinuities is demonstrated by the propagation of a detonation wave. The findings of this study serve as a basis for the development of highly scalable discontinuous Galerkin-based solvers for combustion simulations on massively parallel supercomputers.
\end{abstract}



\end{frontmatter}


\section{Introduction\label{sec:introduction}} \addvspace{10pt}

Direct numerical simulations (DNS) and large-eddy simulations (LES) of reacting flows provide fundamental insights into combustion phenomena \cite{rieth2024,Gruber_CombFlam_2021,Im_CompFlui_2021,Nivarti_PROCI_2017}.
In general, DNS solvers employ higher-order numerical schemes to solve complex problems at extremely high resolutions and are therefore computationally intensive.
Although current state-of-the-art DNS/LES solvers are highly scalable, they encounter communication overheads at extreme scales due to the data synchronization requirements. This issue has been observed in various multi-core/many-core reacting flow solvers \cite{Im_CompFlui_2021,Bialewski_CompFlu_2023,Cant_JCP_2022,Mira_PROCI_2023}, where scalability degrades as the number of processing elements (PEs) increases. Recent studies have proposed methodologies for optimizing overall performance, such as reducing the computational cost of solving chemical kinetics \cite{Im_PROCI_2023}, dimensionality reduction \cite{Malik_PROCI_2024,Dibya_CombFlam_2024}, portable programming models (e.g., Kokkos) \cite{Mira_PROCI_2023} and the introduction of computation-communication overlap at the algorithm level \cite{Schau_PROCI_2023}. While these methods are effective in reducing the computational effort, the fraction of communication cost would appear more prominent, further emphasizing the overhead.

DNS solvers for reacting flows have traditionally been designed with high-order finite-difference methods and are well-suited for handling simple geometries. As the computational capacity of massively parallel systems increases over time, it is becoming increasingly feasible to solve complex combustion problems in practically relevant geometries \cite{Aditya_PROCI_2019,Nivarti_PROCI_2017}.
However, conventional finite-difference methods might not capture subtle flow phenomena in complex geometries because of the difficulty in discretizing irregular boundaries using structured grids.
On the other hand, finite-volume based DNS solvers are more suited for complex geometries but they are restricted to low orders of accuracy and possess less arithmetic intensity.
As a result, they would have to be used at higher resolutions to achieve comparable accuracy as finite-difference methods, which would increase the computational cost.
Furthermore, recent research has focused on developing rotating detonation engines \cite{Wolanski_PROCI_2013,Lu_JPP_2014} and pulse detonation engines \cite{Kailasanath_PROCI_2002} which produce substantial thrust due to the rapid energy release associated with detonations. Therefore, it is of interest to analyze the performance of such engines while accounting for flow discontinuities and detailed chemical kinetics. In this context, the discontinuous Galerkin (DG) method is suitable for designing scalable reacting flow solvers due to its ability to achieve high-order accuracy, handle complex geometries and flow discontinuities, and exhibit high arithmetic intensity \cite{Ihme_JCP_2014}. Parallel DG solvers can be designed to perform point-to-point communications of solutions from the PE boundary elements to the buffer elements of neighboring PEs that are necessary for computing numerical fluxes at the interface.  The algorithm proceeds to the next time step advancement only after all the buffer nodes contain the latest information; this is referred to as the synchronous approach. The explicit synchronization of these communications results in a performance bottleneck at extreme scales, thereby degrading scalability.

To address this issue, asynchrony-tolerant (AT) schemes for finite-difference-based PDE solvers have been developed, and the accuracy and scalability of asynchronous algorithms have been extensively verified \cite{Donzis_JCP_2014,Aditya_JCP_2017,Aditya_arxiv_2019,Komal_JCP_2020,Shubham_JCP_2023,Komal_JCP_2023}. This idea has been extended to the DG method with a recently proposed asynchronous discontinuous Galerkin (ADG) method \cite{Shubham_CMAME_2024} to reduce communication bottlenecks at the mathematical level. This reduction is achieved either by avoiding communication or by relaxing the synchronization between PEs, leading to old/stale data at the PE boundaries.
Asynchrony-tolerant (AT) numerical schemes that use data from multiple time levels were used to approximate the fluxes in PE boundary elements, maintaining the desired order of accuracy.
The order of accuracy of these fluxes was verified by solving simple linear and nonlinear PDEs \cite{Shubham_CMAME_2024}; however, the efficacy of the ADG method in solving problems of high complexity is yet to be investigated.

In this study, we investigate the numerical performance of the ADG method in combustion problems by solving one-dimensional reacting flow equations. Asynchrony-tolerant weighed essentially non-oscillatory (AT-WENO) limiters are derived to handle discontinuities using delayed solutions at PE boundaries. The remainder of this paper is organized as follows. The governing equations are presented in Sec.~\ref{sec:governing_eqns}, and the ADG method and AT WENO limiters and their implementation with simulated delays are explained, respectively, in Sec.~\ref{sec:adg} and Sec.~\ref{sec:algo}. Two test cases, autoignition of lean \ch{H2}–air mixture and propagation of a detonation wave in a \ch{H2}–\ch{O2}–\ch{Ar} mixture, are considered for validation, and their results are discussed in Sec.~\ref{sec:results}. Conclusions are presented in Sec.~\ref{sec:conclusions}.

\section{Governing equations\label{sec:governing_eqns}} \addvspace{10pt}
The one-dimensional equations of conservation of mass, momentum, energy and species are expressed in Eq.~\eqref{eq:ns_vec_eqns}, where $\bm{w}$ is the vector of the conserved variables, $\bm{f}^{c}$ and $\bm{f}^{d}$ are the convective and diffusive fluxes, respectively, and $\bm{s}$ is the source term.
\begin{equation}
\frac{\partial \bm{w}}{\partial t} + \frac{\partial \bm{f}^{c}}{\partial x} = \frac{\partial \bm{f}^{d}}{\partial x} + \bm{s}
\label{eq:ns_vec_eqns}
\end{equation}
%
\begin{equation}
\begin{aligned}
\bm{w} &= 
\left(
    \rho,
    \rho u,
    \rho E,
    \rho Y_1,
    \cdots,
    \rho Y_{N_{s}}
\right)^T\\
\bm{f}^{c} &= 
\left(
    \rho u,
    \rho u^2 + P,
    u(\rho E + P),
    \rho u Y_1,
    \cdots,
    \rho u Y_{N_{s}}
\right)^T\\
\bm{f}^{d} &= 
\left(
    0,
    \tau,
    \tau u - q,
    -\rho Y_1 V_1,
    \cdots,
    -\rho Y_{N_{s}}  V_{N_{s}}
\right)^T\\
\bm{s} &= 
\left(
    0,
    0,
    0,
    W_1 \Dot{\omega}_1,
    \cdots,
    W_{N_{s}} \Dot{\omega}_{N_{s}}
\right)^T
\end{aligned}
\end{equation}
In the vector expressions, $\rho$ is the density, $u$ is the velocity, and $E=u^2/2-P/\rho+h$ is the specific total energy, where $h = \sum_{i=1}^{N_s}Y_ih_i$ is the specific total enthalpy.
$Y_i$ is the mass fraction, $W_i$ is the molecular weight and $\omega_i$ is the molar production rate of species $i$.
$\tau = (4\mu/3)\partial u/\partial x$ is the shear stress, and the heat flux and the species diffusion velocities are
\begin{equation}
q = - k\frac{\partial T}{\partial x} + \sum_{i=1}^{N_s} h_i J_i,~
V_i = -\frac{D_{i}^{mix}}{X_{i}}\frac{\partial X_{i}}{\partial x},
\end{equation}
where $k$ is the thermal conductivity, and $X_i=Y_iW/W_i$, $J_i = \rho Y_i V_i$ and $D_{i}^{mix}$ are the mole fraction, diffusive flux and mixture-averaged diffusion coefficient of species $i$, respectively.

\section{Asynchronous discontinuous Galerkin method\label{sec:adg}} \addvspace{10pt}
In this section, details of the standard discontinuous Galerkin method to solve the governing equations are provided first as a reference. This is followed by a description of the asynchronous discontinuous Galerkin method. Subsequently, new asynchrony-tolerant WENO limiters that are necessary to capture discontinuities are derived.

\subsection{Standard DG method\label{subsec:sdg}} \addvspace{10pt}
The local discontinuous Galerkin method \cite{Du_JCP_2022} is used to solve the governing equations by considering a single flux function $\bm{f}(\bm{w},\bm{q}) = \bm{f}^{c}(\bm{w}) - \bm{f}^{d}(\bm{w},\bm{q})$, where $\bm{q} = \partial \bm{w}/\partial x$ is a vector of the derivatives of the conserved variables.
The computational domain $\Omega = [0,L]$ is divided into a set of $N_E$ non-overlapping elements, denoted by $\Omega \equiv \Omega_h = \bigcup_{e=1}^{N_E} \Omega_e$ where the $e$th element $\Omega_e$ spans $[x_{e},x_{e+1}]$.
Let $\bm{V}_h$ be the finite element space of piecewise smooth functions over the domain $\Omega_h$ and $\bm{V}_{h}^{e}$ be the elemental space spanned by basis polynomial functions $\bm{\phi}_{}^{e}$ of degree up to $N_p$. 
The approximate solution in the element $e$, $\bm{w}_{h}^{e}$, can be expressed as
$\bm{w}_{h}^{e}(x,t) = \sum_{j=0}^{N_p} \bm{\hat{w}}_{j}^{e}(t) \phi_{j}^{e}(x)$,
where $\bm{\hat{w}}_{j}^{e}$ are the degrees of freedom (DoFs) of the element, with $0 \leq j \leq N_p$.
The local solutions are combined to obtain the global solution $\bm{w}_{h} = \bigoplus_{e=1}^{N_E} \bm{w}_{h}^{e}$.
Substituting the approximate solution into Eq.~\eqref{eq:ns_vec_eqns} would result in a residual expressed as
$\bm{\mathcal{R}}_{h}(x,t) = \partial \bm{w}_{h}/\partial t + \partial \bm{f}_{h}/\partial x - \bm{s}_{h}$.
This residual $\bm{\mathcal{R}}_{h}$ has to be minimized to compute the DoFs $\bm{w}_{h}$, which is achieved by making the residual orthogonal to some test functions $\bm{v}_{h}$ in the function space $V_{h}$.
At the element level, the basis polynomials $\bm{\phi}^{e}$ are selected as test functions, and their inner product with the residual is equated to zero.
This results in a system of $N_p+1$ equations to solve for the $N_p+1$ local DoFs of each element, as given by the integral equation in Eq.~\eqref{eq:int_eqn_test}.
\begin{equation}
\begin{split}
    \int\limits_{\Omega_e} \frac{\partial \bm{w}_{h}^{e}}{\partial t} \cdot \bm{\phi}^{e} dx &= \int\limits_{\Omega_e} \bm{f}_{h}^{e}\cdot \frac{d \bm{\phi}^{e}}{dx} dx - \bm{\hat{f}}_{e+1}\cdot\bm{\phi}^{e}(x_{e+1})\\
    &+ \bm{\hat{f}}_{e}\cdot\bm{\phi}^{e}(x_{e}) + \int\limits_{\Omega_e} \bm{s}_{h}^{e}\cdot \bm{\phi}^{e} dx
\end{split}
\label{eq:int_eqn_test}
\end{equation}
A linear transformation is performed from $\Omega_e$ to a reference element $\Omega_R$, where $\xi \in \Omega_R \equiv [-1,1]$, resulting in the semi-discrete weak form,
\begin{equation}
\begin{split}
    \frac{d \bm{w}^{e}_{h}}{dt} = \frac{2}{\Delta x}\bm{\mathcal{M}}^{-1} &\left(\bm{\mathcal{S}}\bm{f}^{e}_h - \bm{\hat{f}}_{e+1}\cdot\bm{\phi}^{e}(x_{e+1})\right. \\
    &\left.+ \bm{\hat{f}}_{e}\cdot\bm{\phi}^{e}(x_{e})\right) + \bm{s}^{e}_{h},
\end{split}
\label{eq:dg_eqn}
\end{equation}
where $\bm{\hat{f}}_e$ and $\bm{\hat{f}}_{e+1}$ denote the numerical fluxes across the interfaces at $x_{e}$ and $x_{e+1}$, respectively.
$\bm{\mathcal{M}}$ and $\bm{\mathcal{S}}$ denote the mass and stiffness matrices, respectively.
\begin{equation}
    [\bm{\mathcal{M}}]_{ij} = \int\limits_{-1}^{1} \phi_{j}(\xi)\phi_{i}(\xi) d\xi,
    [\bm{\mathcal{S}}]_{ij} = \int\limits_{-1}^{1} \phi_{j}(\xi) \frac{d\phi_{i}(\xi)}{d\xi}d\xi
\label{eq:mass_stiff}
\end{equation}
The numerical fluxes are computed using two schemes: the Lax-Friedrichs (LF) flux for the premixed auto-ignition case and the Harten-Lax-van Leer-contact (HLLC) Riemann solver \cite{Toro1994} for the propagation of detonation.
The local Lax-Friedrichs flux across the interface at $x_e$ is expressed as $\bm{\hat{f}}_{e} = (1/2)\left[\bm{f}_{e}^{-}+\bm{f}_{e}^{+}-\lambda\left(\bm{w}_{e}^{+}-\bm{w}_{e}^{-}\right)\right]$,
where $\lambda=\max\{|\bm{f}^{c}_{\bm{w}}(\bm{w}^{+}_{e})|,|\bm{f}^{c}_{\bm{w}}(\bm{w}^{-}_{e})|\}$.
The auxiliary variable is approximated as $\bm{q}_{h}^{e}$ and evaluated similarly using a central flux formulation for $\bm{w}_{h}^{e}$.
The thermodynamic properties and species production rates in Eq.~\eqref{eq:ns_vec_eqns} are computed using Cantera \cite{Cantera}.
Time integration is performed using low-storage explicit Runge-Kutta (LSERK) schemes \cite{Williamson_JCP_1980,Kennedy_ANM_2000}.



\begin{figure}[h!]
    \centering
    \includegraphics[width=8cm,trim=0.5cm 0cm 1cm 0cm,clip]{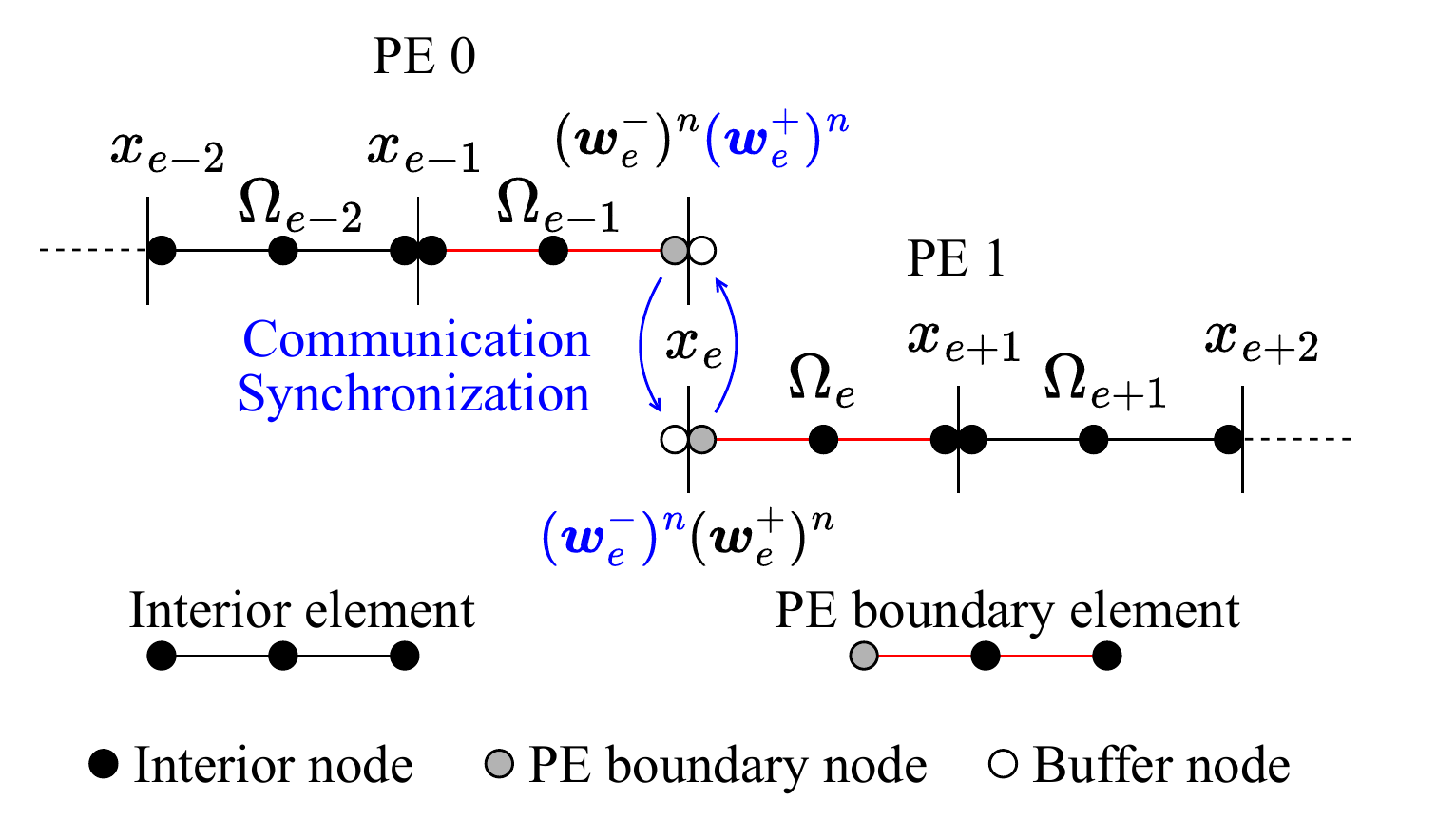}
    \caption{A schematic of synchronized communication at a PE boundary.}
    \label{fig:sync_schematic}
\end{figure}
In parallel implementations of DG, the domain is divided into subdomains, each consisting of interior and PE boundary elements. The fluxes in the interior elements are calculated using standard numerical schemes. However, the computation of fluxes in the boundary elements involves communication of DoFs between PEs. In standard parallel algorithms, the DoFs at the boundary nodes are communicated to the halo/buffer nodes of neighbouring PEs with explicit synchronization, as illustrated in Fig.~\ref{fig:sync_schematic}. The buffer nodes contain the latest information from the neighbouring PEs in this synchronous approach.

\subsection{Asynchronous DG method\label{subsec:atdg}} \addvspace{10pt}
\begin{figure}[h!]
    \centering
    \includegraphics[width=8cm,trim=0.5cm 0cm 1cm 0cm,clip]{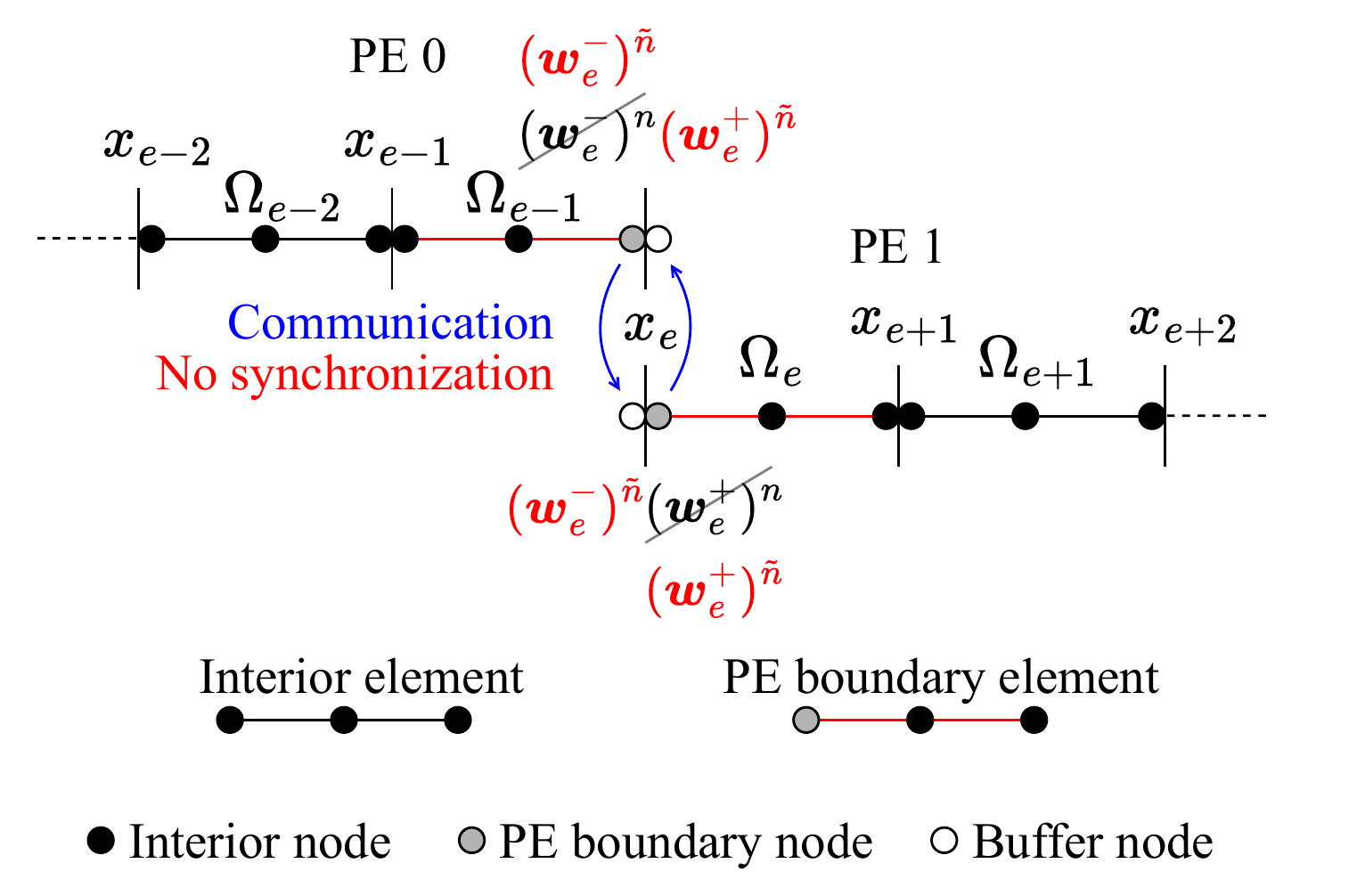}
    \caption{A schematic of communication with relaxed synchronization at a PE boundary.}
    \label{fig:async_schematic}
\end{figure}
In a parallel implementation of the asynchronous discontinuous Galerkin (ADG) method, asynchrony effects are realized either by relaxing synchronization or by avoiding communication between PEs, and the buffer nodes may contain solutions from previous time levels. The extent of relaxation would vary based on a number of factors, such as load balancing among PEs, the communication library, and network topology, resulting in stochastic behavior of delays. In this section, the scenario of relaxed synchronization is considered to describe the ADG method. Let us consider the stencil shown in Fig.~\ref{fig:async_schematic}, where $\Omega_{e-1}$ and $\Omega_{e}$ are the PE boundary elements of PE-0 and PE-1, respectively. The buffer nodes in $\Omega_{e-1}$ and $\Omega_{e}$ may contain delayed data $(\bm{w}_{e}^+)^{\tilde{n}^+}$ and $(\bm{w}_{e}^-)^{\tilde{n}^-}$, respectively, where $\tilde{n}^+$ and $\tilde{n}^-$ denote the stale time levels of data in the respective buffer nodes. The numerical fluxes at $x_e$ in $\Omega_{e-1}$ and $\Omega_e$, respectively, are $\bm{\hat{f}}((\bm{w}_{e}^-)^{n},(\bm{w}_{e}^+)^{\tilde{n}^{+}})$ and $\bm{\hat{f}}((\bm{w}_{e}^-)^{\tilde{n}^{-}},(\bm{w}_{e}^+)^{n})$. To satisfy local conservation at the interface $x_e$, both fluxes must be the same, i.e., the fluxes should be computed using the solutions at a common delayed time level $\tilde{n}=\min(\tilde{n}^+,\tilde{n}^-) = n - \tilde{k}$ with an associated delay $\tilde{k}$. The resulting single-valued flux can be expressed as $\bm{\hat{f}}((\bm{w}_{e}^-)^{\tilde{n}},(\bm{w}_{e}^+)^{\tilde{n}})$ as shown in Fig.~\ref{fig:async_schematic}. The numerical properties of these asynchronous fluxes have been extensively studied and have been verified to be consistent and stable, but result in poor accuracy \cite{Shubham_CMAME_2024}. To overcome this issue, asynchrony-tolerant (AT) fluxes were proposed, where the flux at the PE boundary is approximated using a stencil extended along the temporal direction. The AT flux can be expressed as a combination of the fluxes from previous time levels $n-l$, where $l\in[L_1,L_2]$ with coefficients $\tilde{c}^l$.
\begin{equation}
    \bm{\hat{f}}^{\text{at},n}_{e} \approx \sum_{l=L_1}^{L_2} \Tilde{c}^{l} \bm{\hat{f}}^{n-l}_{e}
    \label{eq:atflux_taylor}
\end{equation}
The desired order of accuracy can be achieved by eliminating the necessary low-order terms in the Taylor series expansion along the temporal dimension.
The size of the temporal stencil depends on the spatial order $(\sim \mathcal{O}(\Delta x^{N_p+1}))$ and the scaling relation in the stability criterion $\Delta t \sim \Delta x^{r}$.
\begin{equation}
    \sum_{l=L_1}^{L_2} \Tilde{c}^{l} \frac{(-l\Delta t)^{\zeta}}{\zeta!} = 
\begin{cases}
    1, & \zeta = 0\\
    0, & 0 < \zeta < (N_p+1)/r
\end{cases}
\label{eq:atflux_constraint}
\end{equation}
Table~\ref{tab:at_coeffs} lists the coefficients $\tilde{c}^{l}$ obtained by solving Eq.~\eqref{eq:atflux_constraint}.
For instance, AT flux that is second order accurate can be obtained as
$\bm{\hat{f}}^{\text{at},n}_{e} = (\tilde{k}+1)\bm{\hat{f}}^{n-\tilde{k}}_{e} - \tilde{k}\bm{\hat{f}}^{n-\tilde{k}-1}_{e}$.

\begin{table}[] \footnotesize
\centering
\caption{Coefficients of fluxes from delayed time levels in Eq.~\eqref{eq:atflux_taylor}.}
\centerline{\begin{tabular}{l l}
\hline
AT order & Coefficients $\tilde{c}^{l}$ with $\tilde{k}=k$                \\ \hline
\rule{0pt}{3ex}
2        & $\tilde{c}^{k} = (k+1), \tilde{c}^{k+1} = -k$                                                                                                                                                                                                \\
\rule{0pt}{6ex}
3        & \begin{tabular}[c]{@{}l@{}}$\tilde{c}^{k} = \left(k^2+3k+2\right)/2$\\ $\tilde{c}^{k+1} = -(k^2+2k)$\\ $\tilde{c}^{k+2} = \left(k^2+k\right)/2$\end{tabular}                                                                                 \\
\rule{0pt}{8ex}
4        & \begin{tabular}[c]{@{}l@{}}$\tilde{c}^{k} = \left(k^3+6k^2+11k+6\right)/6$\\ $\tilde{c}^{k+1} = -\left(k^3+5k^2+6k\right)/6$\\ $\tilde{c}^{k+2} = \left(k^3+4k^2+3k\right)/2$\\ $\tilde{c}^{k+3} = -\left(k^3+3k^2+2k\right)/6$\end{tabular} \\ \hline
\end{tabular}}
\label{tab:at_coeffs}
\end{table}

\subsection{AT-WENO limiters\label{subsec:atweno}} \addvspace{10pt}
Combustion simulations often deal with discontinuities in the flow such as shocks and detonation waves.
Weighted essentially non-oscillatory (WENO) limiters \cite{Zhong_JCP_2013} are typically used to prevent numerical oscillations and capture the discontinuities with minimal numerical dissipation.
The resulting solution in an element is a non-convex weighted combination of the solutions in a stencil containing its neighboring elements.
The flux Jacobian matrix $\bm{A} = \partial \bm{f}^{c}/\partial \bm{w}$ is decomposed as $\bm{A} = \bm{R}\bm{\Lambda}\bm{L}$ where $\bm{\Lambda} = \text{diag}(u-a,u,\dots,u,u+a)$, $\bm{R}$ and $\bm{L}$ are matrices containing right and left eigenvectors of $\bm{A}$ respectively, and $a$ is the speed of sound.
The DoFs of conserved quantities are transformed into the characteristic variables using the relation $\bm{v}_{e+j} = \bm{L}\bm{w}_{e+j}$, where $j=-1,0,1$ and $\bm{L} = \bm{L}(\bm{\bar{w}}_{e})$.
Troubled cells are identified using total variation diminishing (TVD) slope limiters as detailed in \cite{Zhong_JCP_2013}, and the limiting procedure is applied to the characteristic variables as given in Eq.~\eqref{eq:weno_limiter}.
\begin{equation}
    \bm{v}_{e}^{\text{lim}} = \omega_{-1}\bm{\tilde{v}}_{e-1} + \omega_{0}\bm{v}_{e} + \omega_{1}\bm{\Tilde{v}}_{e+1},
    \label{eq:weno_limiter}
\end{equation}
where $\bm{\tilde{v}}_{e\pm1} = \bm{v}_{e\pm1} - \bm{\bar{v}}_{e\pm1} + \bm{\bar{v}}_{e}$.
The non-linear weights $\omega_{j}$ are
\begin{equation}
\begin{aligned}
\omega_j &= \frac{\bar{\omega}_j}{\sum_{s}\bar{\omega}_s},\ 
\bar{\omega}_j = \frac{\gamma_j}{(\epsilon+\beta_j)^2} \\
\beta_j &= \sum_{s=1}^{N_p} \int\limits_{\Omega_e} (\Delta x)^{2s-1} \left(\frac{\partial^s}{\partial x^s}\bm{v}_{e+j}\right)^2 dx,
\end{aligned}
\end{equation}
where $j=-1,0,1$ and $\sum_{s}\omega_s = 1$.
$\gamma_j$ are the linear weights, $\epsilon$ is a small number taken as $10^{-6}$ to avoid the possibility of division by zero, and $\beta_j$ is the smoothness indicator which depends on the derivatives of the solution.
After the limiting procedure, the conserved variables are obtained using the transformation $\bm{w}^{\text{lim}}_{e} = \bm{R}\bm{v}^{\text{lim}}_{e}$, where $\bm{R} = \bm{R}(\bm{\bar{w}}_{e})$.
The standard WENO limiters are implemented in parallel solvers by communicating the DoFs from PE boundary elements to buffer elements of neighbouring PEs after every stage of the time integration. 

\begin{figure}
    \centering
    \includegraphics[width=8cm,trim=0cm 0cm 7cm 0cm,clip]{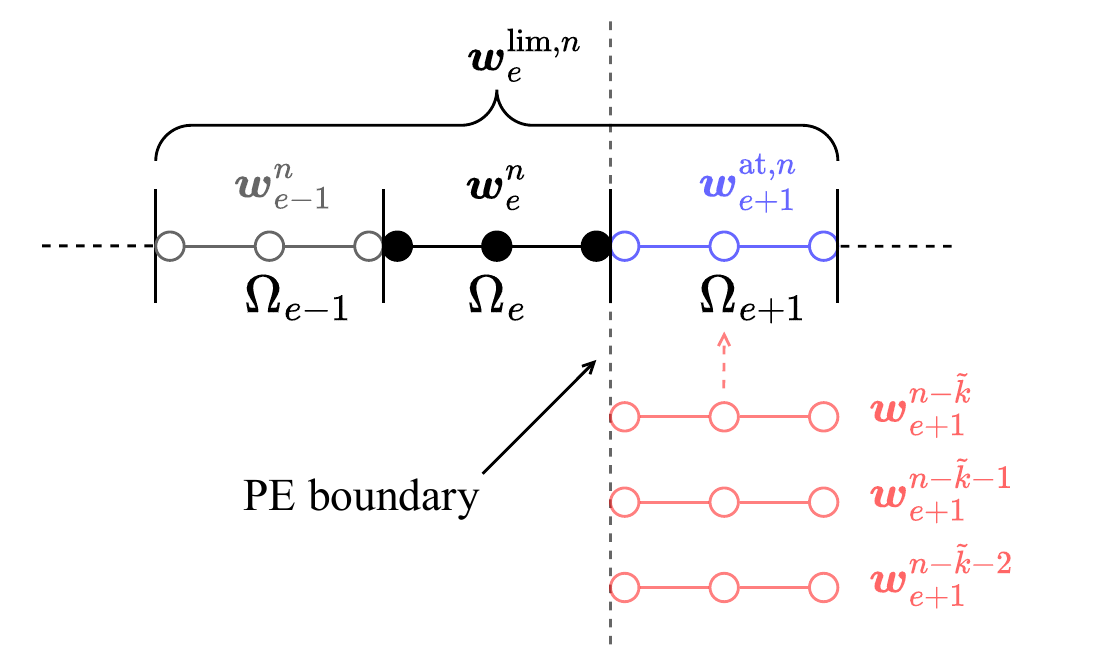}
    \caption{A schematic of the asynchrony-tolerant (AT) WENO stencil at a right PE boundary element.}
    \label{fig:at_weno}
\end{figure}

In the presence of asynchrony, the WENO stencil of a PE boundary element may not contain the latest updated DoFs from the neighbouring PE.
Applying the limiter with this stencil can cause a loss of accuracy, particularly if the solution is discontinuous near the PE boundary.
To overcome this, a modified stencil is proposed, where the DoFs of the element in the neighbouring PE are approximated using data from previous time levels.
The size of this stencil and the coefficients are again dictated by the required order of accuracy and the stability scaling relation, as discussed in Sec.~\ref{subsec:atdg}.
Consider a stencil centered at $\Omega_e$ which is the right boundary element of a PE, as shown in Fig.~\ref{fig:at_weno}.
The DoFs in $\Omega_{e+1}$ are approximated using AT schemes as given in Eq.~\eqref{eq:at_scheme_w} and are used in the candidate stencil for the limiting procedure in $\Omega_e$.
Thus, the AT-WENO limiter expression can be obtained by modifying the traditional WENO procedure to the form $\bm{v}_{e}^{\text{lim},n} = \omega_{-1}\bm{\tilde{v}}^{n}_{e-1} + \omega_{0}\bm{v}^{n}_{e} + \omega_{1}\bm{\Tilde{v}}^{\text{at},n}_{e+1}$.
\begin{equation}
    \bm{w}^{\text{at},n}_{e+1} \approx \sum_{l=L_1}^{L_2} \Tilde{c}^{l} \bm{w}^{n-l}_{e+1}
    \label{eq:at_scheme_w}
\end{equation}
%
%
The temporal stencil used for approximating the buffer element DoFs can be generalized to an arbitrary spatio-temporal stencil where the DoFs in the buffer element are updated using delayed values at other nodes in the element.
This hybrid stencil could potentially further reduce the communication volume using buffers with less time levels, achieving the same order of accuracy.

\section{Implementation\label{sec:algo}} \addvspace{10pt}
The implementation of the standard DG method discussed in Sec.~\ref{subsec:sdg} will be referred to as the synchronous algorithm (SA) in subsequent sections.
The ADG method is implemented using two approaches: the communication avoiding algorithm (CAA) and the synchronization avoiding algorithm (SAA) \cite{Komal_JCP_2020,Shubham_JCP_2023}.
These algorithms are implemented on a serial program, emulating the corresponding parallel algorithms that can be implemented using a communication library like Message Passing Interface (MPI).
In this serial solver, the computational domain is divided into $P$ subdomains mapped to as many processing elements.
Asynchrony is introduced at PE boundaries using simulated delays which are bounded by a maximum allowable delay $L$.
It was mathematically shown that the truncation error in the AT fluxes depends on simulation parameters such as the number of PEs and delay statistics \cite{Shubham_CMAME_2024}.
Therefore, it is important to assess the effect of these parameters on the numerical errors incurred in the simulations.
For these reasons, the serial implementation with simulated delays is considered, as it provides us complete control over the delay statistics and aids the analysis of the ADG method across different configurations.

\subsection{Communication avoiding algorithm (CAA)\label{subsec:caa_algo}} \addvspace{10pt}
In CAA, delays are introduced by avoiding communication between PEs over a predetermined number of steps in the time advancement procedure.
These delays are considered to be uniform across PEs and exhibit periodic behaviour over time.
This deterministic delay is simulated by allowing the algorithm to proceed using fluxes from delayed time levels.
Let $n$ be the current time step in the time advancement procedure.
If $n\%L\neq0$, the delay is incremented by one level ($\tilde{k}++$) and the algorithm moves to the next step with the updated delay.
In a parallel implementation of CAA, this would be achieved by skipping the communication requests such as \texttt{MPI\_Send} and \texttt{MPI\_Recv} over $L$ time steps, including the intermediate RK stages, if multi-stage RK schemes are used \cite{Shubham_JCP_2023}.
If $n\%L=0$, the maximum allowable delay is attained and the delay is reset ($\tilde{k}=0$).
With MPI, the solutions at PE boundary nodes are communicated to buffer nodes and synchronized under this condition and the delay is reset.

\subsection{Synchronization avoiding algorithm (SAA)\label{subsec:saa_algo}} \addvspace{10pt}
In SAA, the synchronization of communications between PEs is relaxed and the delays with random behaviour at PE boundaries, as discussed in Sec.~\ref{subsec:atdg}.
In a parallel implementation, non-blocking requests such as \texttt{MPI\_Isend} and \texttt{MPI\_Irecv} are initiated at every time step and the algorithm is allowed to proceed without explicit synchronization calls such as \texttt{MPI\_Wait}.
The status of existing requests is obtained using \texttt{MPI\_Test}, and if the delay exceeds $L$, the requests are synchronized and the delay is reset.
It should be noted that additional overhead should be imposed to ensure that local conservation is maintained using a common delayed time level, as discussed in Sec.~\ref{subsec:atdg}.
The communication delays at the PE boundaries are simulated in the serial program using a probability distribution and a random number generator.
For a given $L$, a probability distribution $\{p_0,p_1,\cdots,p_{L-1}\}$ is considered over the entire simulation such that $\sum_{k=0}^{L-1} p_k = 1$, where $p_k$ denotes the probability that $\tilde{k}=k\in\{0,1,2,\cdots,L-1\}$.
Based on the probability set, the interval $[0,1]$ is partitioned into $L$ successive bins.
A number between 0 and 1 is drawn randomly and is matched with one of the bins, and the delay corresponding to the mapped bin is considered.
This procedure, adopted from \cite{Shubham_CMAME_2024}, is carried out at each PE boundary and at each time step.
Different probability distributions are considered based on the probability of delays observed in parallel implementations of SAA in recent works \cite{Komal_JCP_2020,Shubham_JCP_2023}.
The probability sets of delays considered in this study are: set-1: $\{0.3,0.4,0.3\}$, set-2: $\{0.5,0.3,0.2\}$ and set-3: $\{0.8,0.1,0.1\}$.


\section{Results\label{sec:results}} \addvspace{10pt}
Two test cases are considered in this study to assess the numerical accuracy of the ADG method: premixed auto-ignition of \ch{H2}-air mixture, and propagation of a detonation wave.
The premixed auto-ignition problem is considered because the ignition delay time associated with the mixture is highly sensitive to the evolution of minor species involved in the detailed chemistry.
Therefore, the effect of numerical errors introduced at PE boundaries due to communication delays on the mass fraction of these intermediate species, and in turn on the progression of the flame front, needs to be studied.
The detonation wave problem is considered because practical combustion applications involve supersonic flows with complex discontinuities such as shocks and detonations, for instance in the context of detonation engines discussed in Sec.~\ref{sec:introduction}.
Therefore, it is important to assess the efficacy of the ADG method and the AT-WENO limiters derived in Sec.~\ref{subsec:atweno} in accurately capturing discontinuities in the solution.

\subsection{Auto-ignition of premixed \ch{H2}-air\label{subsec:auto_ign}} \addvspace{10pt}
The spontaneous ignition of a premixed fuel-air mixture at constant volume is a canonical problem of relevance in internal combustion engines.
A lean \ch{H2}-air mixture with equivalence ratio $0.4895$ is considered in a one-dimensional domain with an initial temperature of $600~\text{K}$.
A temperature hot spot represented by a Gaussian profile is introduced at the center with a peak temperature of $1200~\text{K}$.
Periodic boundary conditions are imposed on the physical boundaries to emulate constant volume combustion.
A detailed reaction mechanism involving 9 species and 21 reactions is used to solve the stiff chemical kinetics \cite{Li_IJCK_2004}.
For the asynchronous algorithms, the domain is divided into 8 PEs, and a maximum delay of 3 is considered.
After an induction period during which intermediate species are formed at the hot spot, the mixture ignites spontaneously, accompanied by a rapid increase in temperature.
The ignition at the center is followed by the emergence of two ignition fronts, moving towards the left and right boundaries, away from the hot spot.
The instantaneous profiles of different flow quantities in Fig.~\ref{fig:autoign_profiles} show that the asynchronous algorithms are in excellent agreement with the synchronous algorithm in capturing the flame front.
In addition, energy spectra of $T$ and $Y_{HO_2}$ are shown in Fig.~\ref{fig:autoign_profiles}, demonstrating the good agreement of AT fluxes with the standard fluxes at all wavenumbers ($\kappa$).
The formal order of accuracy of the ADG method is verified by simulating the problem at different resolutions and comparing the solutions to that obtained at a higher resolution with $N_E=1024$ and $N_p=3$.
The ensemble average error $\langle \bar{E}\rangle$ is obtained by averaging over all elements and multiple independent simulations to take into account the randomness in delays.
The convergence plot for velocity is shown in Fig.~\ref{fig:autoign_order}(a) for second and third orders of accuracy.
In addition, the error scales with the simulation parameters as
$\langle \bar{E} \rangle \sim P \Delta t^{N_p+1} \sum_{m=1}^{N_p+1}\gamma_{m}\overline{\tilde{k}^{m}}/N_E$
where $\gamma_m$ denotes the coefficient of $m$th moment of delay \cite{Shubham_CMAME_2024}.
The linear scaling of the errors with the number of PEs ($P$) is demonstrated in Fig.~\ref{fig:autoign_order}(b) and agrees with theoretical predictions.

\setlength{\unitlength}{1cm}
\begin{figure}[h!]
    \centering
    \includegraphics[width=6cm,trim=0cm 0cm 0cm 0cm,clip]{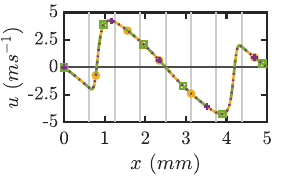}
    \begin{picture}(0,0)
    \put(-4.7,\yloc){(a)}
    \end{picture}\hspace{-0.35cm}
    \includegraphics[width=6cm,trim=0cm 0cm 0cm 0cm,clip]{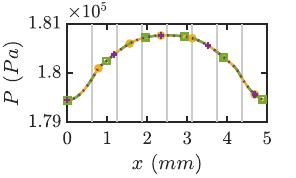}
    \begin{picture}(0,0)
    \put(-4.7,\yloc){(b)}
    \end{picture}\hspace{-0.35cm}
    \includegraphics[width=6cm,trim=0cm 0cm 0cm 0cm,clip]{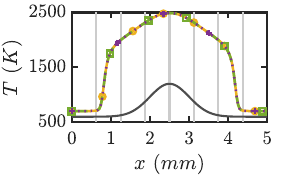}
    \begin{picture}(0,0)
    \put(-4.5,\yloc){(c)}
    \end{picture}\hspace{-0.35cm}
    \includegraphics[width=6cm,trim=0cm 0cm 0cm 0cm,clip]{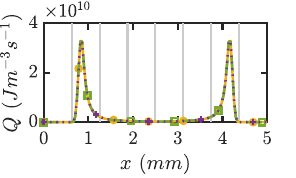}
    \begin{picture}(0,0)
    \put(-5.1,\yloc){(d)}
    \end{picture}\hspace{-0.35cm}
    \includegraphics[width=6cm,trim=0cm 0cm 0cm 0cm,clip]{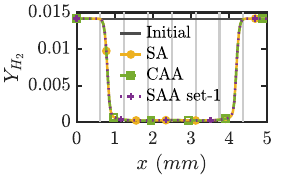}
    \begin{picture}(0,0)
    \put(-4.5,\yloc){(e)}
    \end{picture}\hspace{-0.35cm}
    \includegraphics[width=6cm,trim=0cm 0cm 0cm 0cm,clip]{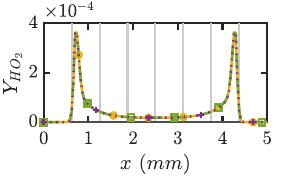}
    \begin{picture}(0,0)
    \put(-5.2,\yloc){(f)}
    \end{picture}\hspace{-0.35cm}
    \includegraphics[width=6cm,trim=0cm 0cm 0cm 0cm,clip]{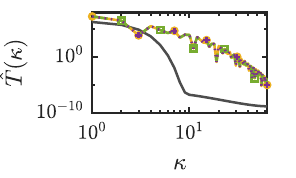}
    \begin{picture}(0,0)
    \put(-4.1,\yloc){(g)}
    \end{picture}\hspace{-0.35cm}
    \includegraphics[width=6cm,trim=0cm 0cm 0cm 0cm,clip]{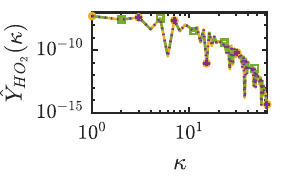}
    \begin{picture}(0,0)
    \put(-4.1,\yloc){(h)}
    \end{picture}
    \caption{Solution profiles ((a)-(f)) and spectra ((g), (h)) of auto-ignition with DG(1)-RK2 and ADG(1)-AT2-RK2 with $N_E=128$, $P=8$ and $L=3$ at $t=1.4\times10^{-4} s$. Grey vertical lines denote processing element (PE) boundaries.}
    \label{fig:autoign_profiles}
\end{figure}
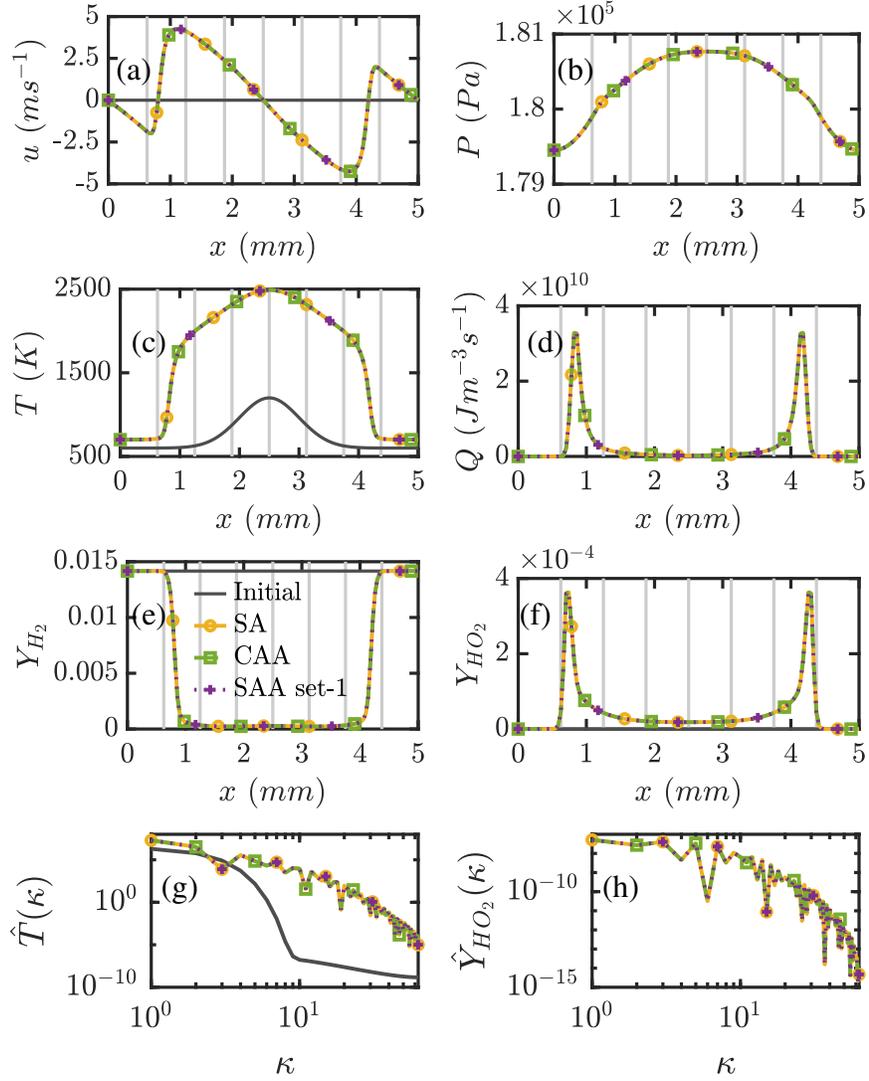

\begin{figure}
    \centering
    \includegraphics[width=6cm,trim=0cm 0cm 0cm 0cm,clip]{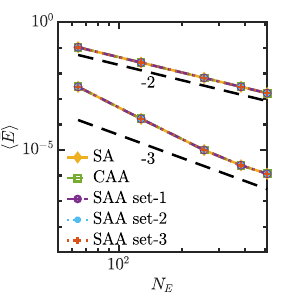}
    \begin{picture}(0,0)
    \put(-1.4,1.2){(a)}
    \end{picture}
    \includegraphics[width=6cm,trim=0cm 0cm 0cm 0cm,clip]{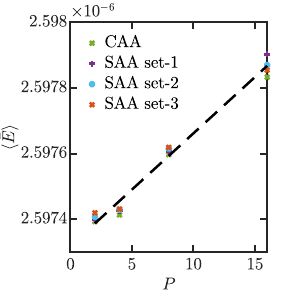}
    \begin{picture}(0,0)
    \put(-1.4,1.2){(b)}
    \end{picture}
    \caption{(a) Convergence plot of $u$ for multiple orders of accuracy, and (b) scaling of $Y_{H_2}$ error with number of processing elements for $N_p=1$, $N_E=256$, for premixed auto-ignition.}
    \label{fig:autoign_order}
\end{figure}


\subsection{One-dimensional detonation wave\label{subsec:deton}} 
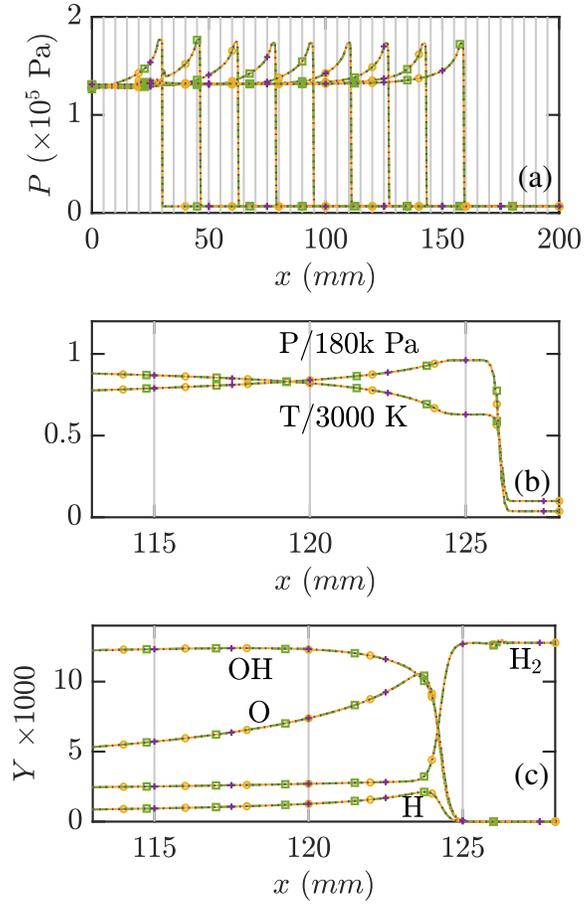
\begin{figure}
    \centering
    \includegraphics[width=8cm,trim=0cm 0cm 0cm 0cm,clip]{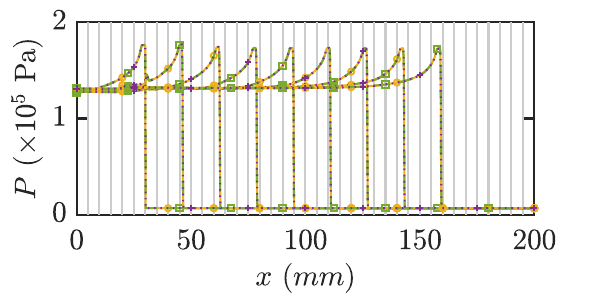}
    \begin{picture}(0,0)
    \put(6.7,5.5){(a)}
    \end{picture}
    \centering
    \includegraphics[width=8cm,trim=0cm 0cm 0cm 0cm,clip]{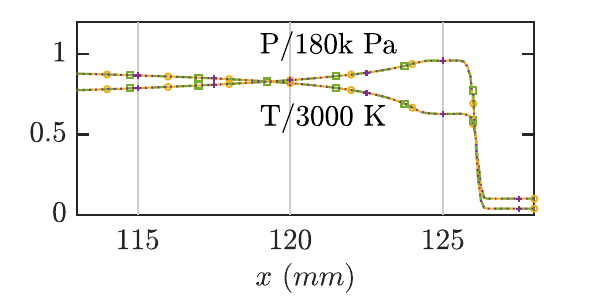}
    \begin{picture}(0,0)
    \put(6.7,5.5){(b)}
    \end{picture}
    \centering
    \includegraphics[width=8cm,trim=0cm 0cm 0cm 0cm,clip]{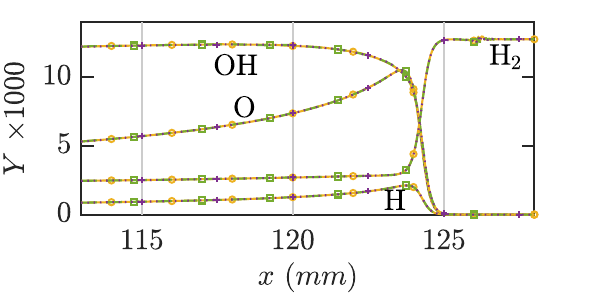}
    \begin{picture}(0,0)
    \put(-1.4,1.6){(c)}
    \end{picture}
    \caption{(a) Time evolution of pressure. (b) Pressure and temperature, and (c) mass fractions of select species at a specific time instant. Simulation parameters $N_E=2000$, $N_p=2$, $P=40$ and $L=3$. Grey vertical lines denote PE boundaries.}
    \label{fig:detonation_profiles}
\end{figure}

\addvspace{10pt}
The numerical accuracy of the ADG method and AT-WENO limiters is evaluated by solving a one-dimensional detonation problem that has been extensively studied \cite{Deiterding2003,Houim_JCP_2011,Ihme_JCP_2014,Komal_JCP_2023}.
A quiescent medium of stoichiometric \ch{H2}/\ch{O2}/\ch{Ar} mixture with the molar ratio $\ch{H2}:\ch{O2}:\ch{Ar}=2:1:7$ is initially at pressure $6670~\text{Pa}$ and temperature $298~\text{K}$.
A normal shock wave corresponding to the \textit{Chapman-Jouguet} (CJ) condition is initialized near the left boundary using SDToolbox \cite{Kao_SDT2024}.
A detailed mechanism involving 9 species and 34 reactions \cite{WestbrookCombFlam_1982} is used to model the chemistry.
Supersonic outflow boundary conditions are imposed at both the left and right boundaries.
The domain of length $200~\text{mm}$ is discretized into 2000 elements at a resolution of $100~\mathrm{\mu m}$ and solved with $N_p=2$.
For the asynchronous algorithms, the domain is divided into 40 PEs and a maximum allowable delay of 3 is considered.
Standard WENO limiters are used with SA while AT-WENO limiters are used with CAA and SAA.
The mixture is compressed by the leading shock followed by an induction region (also called the \textit{von Neumann} state) and a reaction zone.
Figure~\ref{fig:detonation_profiles}(a) shows the pressure at different time instants as the detonation propagates.
The temperature, pressure and mass fractions of a few species at a specific time instant are plotted in Fig.~\ref{fig:detonation_profiles}(b) and~\ref{fig:detonation_profiles}(c).
The converged values of the von Neumann pressure and temperature and the detonation speed are reported in Table~\ref{tab:detonation_vals} for the three algorithms.
The results obtained using the ADG method and AT-WENO limiters are in excellent agreement with the standard DG method and WENO limiters. These results are also similar to those reported in previous studies.

\begin{table}[h!] \footnotesize
\centering
\caption{Target quantities of detonation wave propagation for the three algorithms.}
\centerline{\begin{tabular}{l l l l}
\hline
Algorithm & $p_{\text{vN}}$ (kPa) & $T_{\text{vN}}$ (K) & $u_{\text{det}}$ (m/s) \\ \hline
SA         & 173.33 & 1889.8 & 1610.0  \\
CAA        & 173.37 & 1890.5 & 1610.0  \\
SAA set-1  & 173.35 & 1890.7 & 1610.4  \\ \hline
\end{tabular}}
\label{tab:detonation_vals}
\end{table}

\section{Conclusions\label{sec:conclusions}} \addvspace{10pt}
Discontinuous Galerkin (DG) method has been widely considered in recent times to develop scalable solvers for combustion simulations for its many desirable numerical properties. Unfortunately, such parallel implementations of DG incur communication overheads at extreme scales, adversely affecting scalability. In this study, the application of the recently proposed asynchronous discontinuous Galerkin (ADG) method is extended to complex reacting flows. The synchronization requirements at the processor boundaries are relaxed by introducing communication delays at the boundary elements. To maintain the desired order of accuracy, asynchrony-tolerant (AT) fluxes are approximated using values from multiple delayed time levels at the processing element (PE) boundaries. For flows with discontinuities, standard WENO limiters would result in poor accuracy when they are used with delayed data at PE boundaries. To overcome this issue, new asynchrony-tolerant WENO (AT-WENO) limiters are derived using AT schemes to approximate the buffer elements at PE boundaries. The ADG method is implemented on a serial solver using two algorithms that avoid communication and relax synchronization with simulated delays. The error convergence of the ADG method is verified on spontaneous ignition of a fuel lean mixture and the numerical accuracy of the new AT-WENO limiters is verified on the propagation of a detonation. The results of this manuscript provide a foundation for developing a highly scalable DG-based solver for reacting flows using the ADG method.
\section*{Acknowledgements}
The authors thank Dr. Shubham Kumar Goswami and Aparna Reniguntla for their inputs to the code.



\bibliographystyle{main-style}
\bibliography{main}





\end{document}